\documentclass{moriond}

\usepackage{subfig}
\usepackage{hyperref}

\newcommand{\be}{\begin{eqnarray}}

\newcommand{\ee}{\end{eqnarray}}

\begin{document}
\vspace*{1cm}

\title{Searching for oscillations in the primordial power spectrum}

\author{P.~Daniel Meerburg} 
\address{Department of Astrophysical Sciences, Princeton University, Princeton, NJ 08540 USA. }
\author{David N. Spergel}
\address{Department of Astrophysical Sciences, Princeton University, Princeton, NJ 08540 USA. }
\author{ Benjamin D.~Wandelt}
\address{CNRS-UPMC Univ. Paris 06, UMR7095, Institut d`Astrophysique de Paris, 98bis Bd. Arago, F-75014, Paris, France}
\address{Departments of Physics and Astronomy, University of Illinois at Urbana-Champaign, Urbana, IL 61801, USA}

\maketitle
\abstracts{A small deviation from scale invariance in the form of oscillations in the primordial correlation spectra has been predicted by various cosmological models. In this paper we review a recently developed method to search for these features in the data in a more effective way. By Taylor expanding the small features around the `background' cosmology, we have shown we are able to improve the search for these features compared to previous analyses. In this short paper we will extend that work by combining this method with a multi nested sampler. We recover our previous findings and are able to do so in 192 CPU hours. We will also briefly discuss the possibility of a long wavelength feature in the data to alleviate some tension between the data and the $\Lambda$CDM + $r$ concordance cosmology. 
 }

\section{Introduction}

The Planck data provided unprecedented detail of the small
fluctuations in the cosmic microwave background (CMB) \cite{PlanckInflation2013}.  These results confirm WMAPÕs conclusions that the fluctuations are adiabatic, Gaussian and almost scale invariant \cite{2003ApJS..148..175S}. This constrains models and model parameters of the early
Universe, but despite the quality of the data they remain inconclusive
of the details. In particular, small deviations from scale invariance
are still permitted, and so are small levels of non-Gaussianity. In
recent years models have been proposed which predict oscillatory
features on top of an almost scale invariant spectrum \cite{2005tsra.conf....1G,2010JCAP...06..009F,2011JCAP...01..030A,2013JCAP...03..004D}. The amplitude
and frequency of these oscillations is usually associated with
fundamental parameters of the model, be that the axion decay constant
in axion-monodromy inflation, or the energy scale at which particles
are injected into the free vacuum during inflation in more exotic
models. 

This proceedings is set up as follows. In Section \ref{sec:methodology} we will briefly review a recently proposed method to effectively search for small, vastly oscillating signatures in the CMB power spectrum \cite{Meerburg2014a,Meerburg2014b}. The method is build upon the observation that these features have a small amplitude. Hence, it is possible to significantly speed up the calculation of the time consuming transfer functions through a Taylor expansion. We show that our
method recovers oscillations in Planck-like simulated data at a few
times $10^{-2}$ level. Applying our analysis to Planck and WMAP confirms that this method recovers previously obtained results and improves on existing constraints set by the Planck team \cite{PlanckInflation2013}. We have recently combined our method with the nested sampler MULTINEST \cite{Feroz2009,Feroz2013}, which allows us to recover the posterior distribution of the frequency parameter in the Planck data over a large range of frequencies in about 
192 CPU hours. We will compare the results of MULTINEST with our previous work in Section \ref{sec:multinest}. In Section \ref{sec:BICEP} we will briefly comment on oscillations in light of the BICEP2 results \cite{BICEP2014} and will conclude in Section \ref{sec:conclusions}.

\section{Background and Methodology}\label{sec:methodology}

In this section we will review our developed method in \cite{Meerburg2014a} to search for resonant features applied to the recently released  Planck CMB data \cite{Meerburg2014b}. 
We consider two distinct theoretically motivated models:

\be
_1\Delta^2_{\mathcal{R}}(k)  = A_1 \left(\frac{k}{k_*} \right)^{m}\left(1+A_2 \cos [\omega_1 \log k/k_* +\phi_1]\right),
\label{eq:powerspectra1}
\ee
\be
_2\Delta^2_{\mathcal{R}}(k) = B_1\left(\frac{k}{k_*} \right)^{m}\left(1+B_2 k^n \cos [\omega_2 k +\phi_2]\right).
\label{eq:powerspectra2}
\ee
We refer to the first model as the ``log-spaced oscillations model" and the second model as the ``linear oscillations model".
For example, axion-monodromy inflation produces features that can be described by the logarithmic oscillations model with  $A_1 = H^2/(8\pi^2\epsilon)$, $m=n_s-1$, $A_2 = \delta n_s$, $\omega_1 = -(\phi_*)^{-1}$ and $\phi_1 = \phi_*$. Model that include the effects from a possible boundary on effective field theory (BEFT) predict features that can be described by the linear oscillations model with $B_1 = H^2/(8\pi^2\epsilon)$, $m=n_s-1$, $B_2  = \beta/a_0M$, $n=1$, $\omega_2= 2/a_0 H$ and $\phi_2=\pi/2$. Both initial state modifications and multiverse models \cite{2013JCAP...03..004D} can produce logarithmic oscillations, while sharp features generate a power spectrum with linear oscillations (although the amplitude is typically damped as a function of scale). Constraints on oscillations in the WMAP CMB data have been attempted in e.g. Refs.~\cite{2004PhRvD..69h3515M,2011PhRvD..84f3515D,2007PhRvD..76b3503H,2009PhRvD..79h3503P,Meerburg2012,Flauger2013,2013PhRvD..87h3526A,2014PhRvD..89j3006A}. 

The method we develop relies on the observation that the features has a small amplitude and therefore we can write the final spectrum as 
\be
C_{\ell} & = & C_{\ell}^u + C_{\ell}^p,
\ee
where $C_{\ell}^u$ is the ``unperturbed part" (i.e. without modulations) and $C_{\ell}^p$ is the perturbed part (i.e. with oscillations). The unperturbed part is computed in the usual way, while we can expand the perturbed part around the best-fit parameter values $\bar{\Theta}$ (containing all the plain vanilla $\Lambda$CDM parameters $\Omega_b h^2$, $\Omega_{cdm}h^2$, $\tau$, $A_s$, $n_s$ and $H_0$), i.e., 
\be
C_{\ell}^p(\omega,\phi, A,\Theta) & = & {\color{black} \bar{C}_{\ell}^{p(\alpha)} +\bar{C}_{\ell}^{p(\beta)}+ \sum (\Theta_i-\bar{\Theta}_i)( \bar{C}_{\ell,\Theta_i}^{p(\alpha)}+ \bar{C}_{\ell,\Theta_i}^{p(\beta)}) + \mathcal{O}((\alpha+\beta)\Theta_i^2)}.
\ee
The superscript $\alpha$ and $\beta$ refer the the phase of the primordial signal in Eq. \ref{eq:powerspectra1} and Eq. \ref{eq:powerspectra2}, which can easily be rewritten as renormalized amplitudes of a sum of cosine and sine functions. The advantage of writing the spectrum in this form is clear: for rapidly oscillations functions, one needs to set high resolution both in $k$ and in $\ell$, hence computing the spectra is time consuming, taking of the order of several minutes on a single CPU for the highest oscillations. Through this expansion one can precompute the $ \bar{C}$, $ \bar{C}_{\ell,\Theta_i}$ and higher order derivatives, and store them for a large number of frequencies (writing it in the form above, the phase and amplitude can be altered by wighted sums). As a result, one can now perform a simple Metropolis-Hasting Markov Chain Monte Carlo (MH MCMC) using a modified version of COSMOMC \cite{Lewis2002} for a given frequency and obtain results relatively fast, while allowing all parameters to vary continuously. Higher accuracy is obtained by expanding to higher order in the $\Theta$ parameters. 

\begin{figure}
 
  \centering
 
  \subfloat[Improvement of fit versus $\omega_1$ for several input amplitude's. $A_2=0.1$ and $0.05$ are recovered, while $A_2=0.01$ is not. The `oscillations' are a consequence of the noise (which is the same for all 3 simulations). It is clear that features in the noise can amplify and de-amplify some of the signal. ]{\label{fig:simlog100}\includegraphics[width=0.45\textwidth]{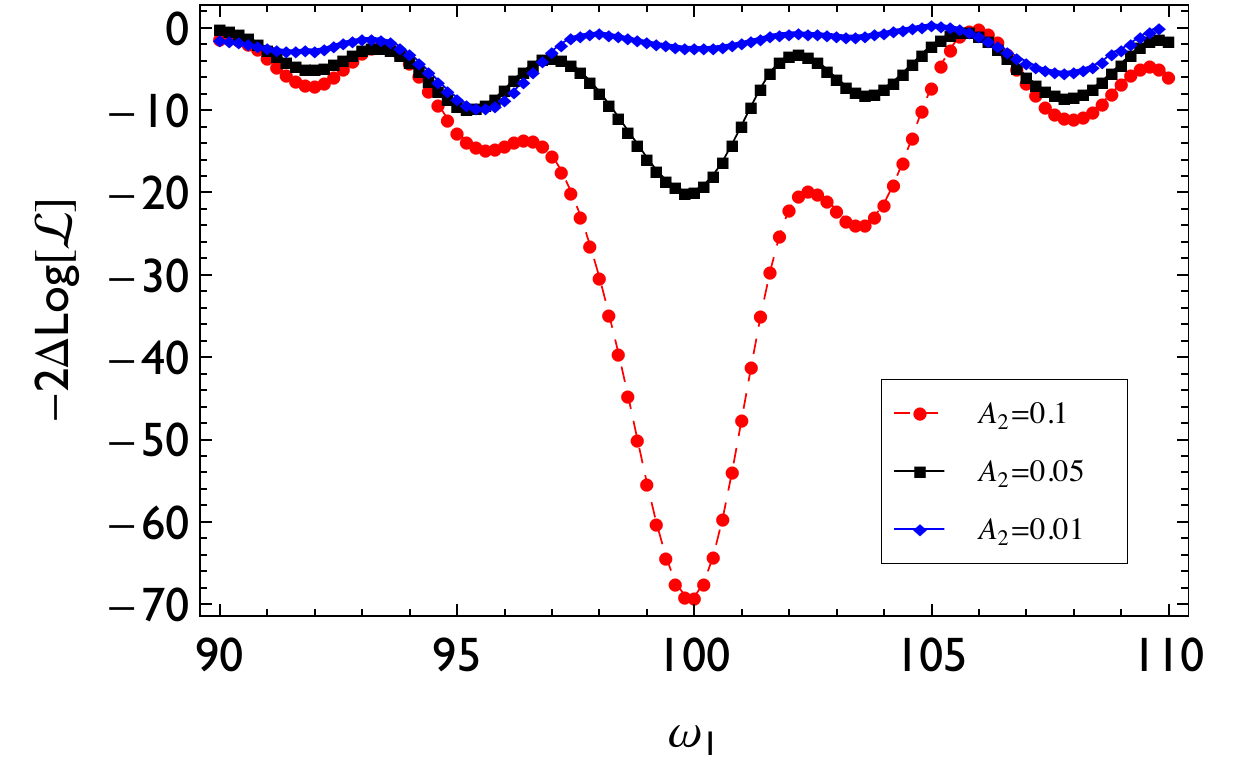}} \qquad
  \subfloat[Results for the best-fit (black) and marginalized frequency distribution (red). Compare to top of Fig. 19 in the Planck inflation paper. ]{\label{fig:multinest}\includegraphics[width=0.45\textwidth]{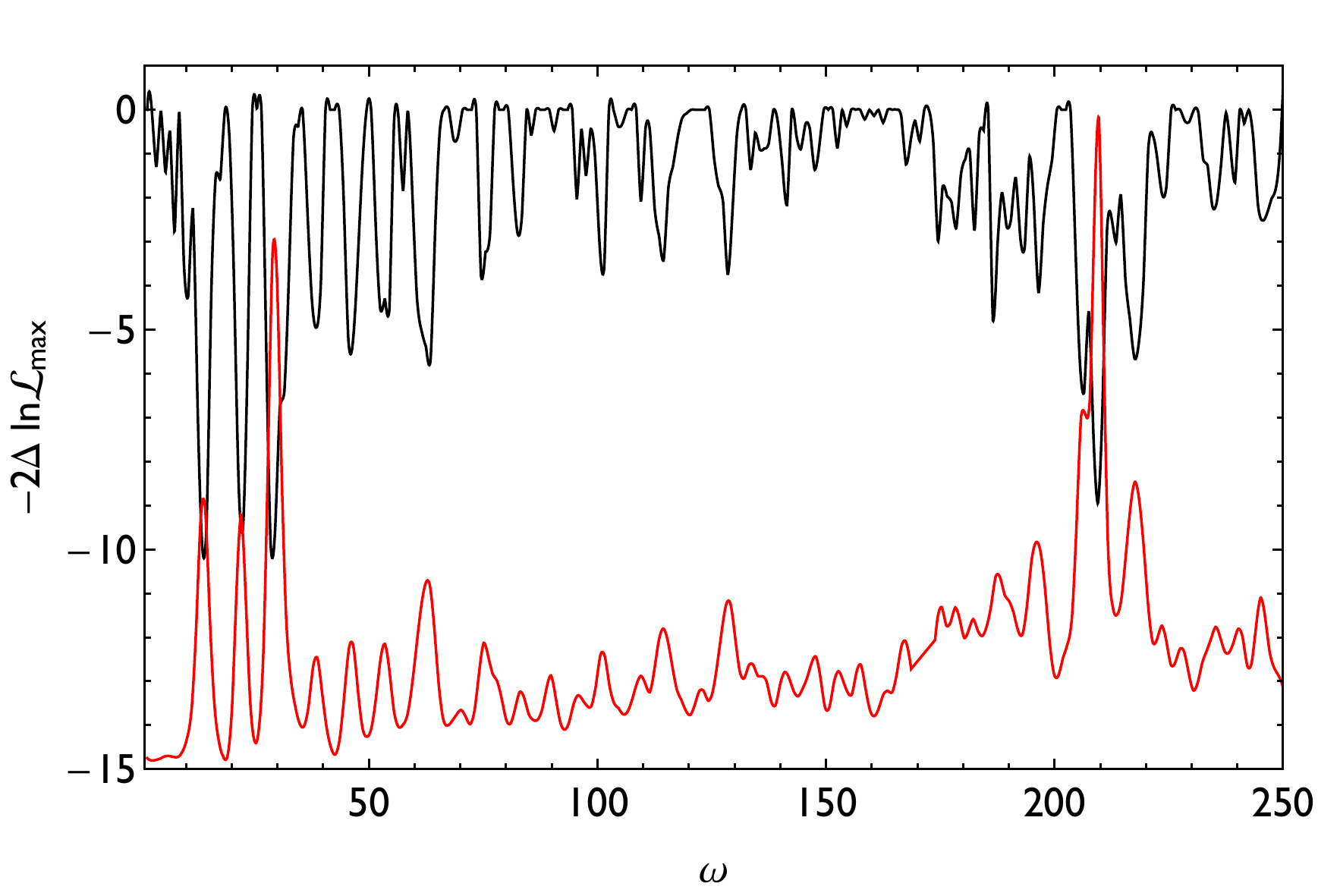}}
    \caption{}
  \label{fig:test}
\end{figure}

We investigated the validity of this expansion in \cite{Meerburg2014a} by generating mock Planck data, and applying this method to recover the inserted signal. An example of such a search and its results in shown in Fig. \ref{fig:simlog100}. 

\begin{figure}[htbp] 
   \centering
   \includegraphics[width=6in]{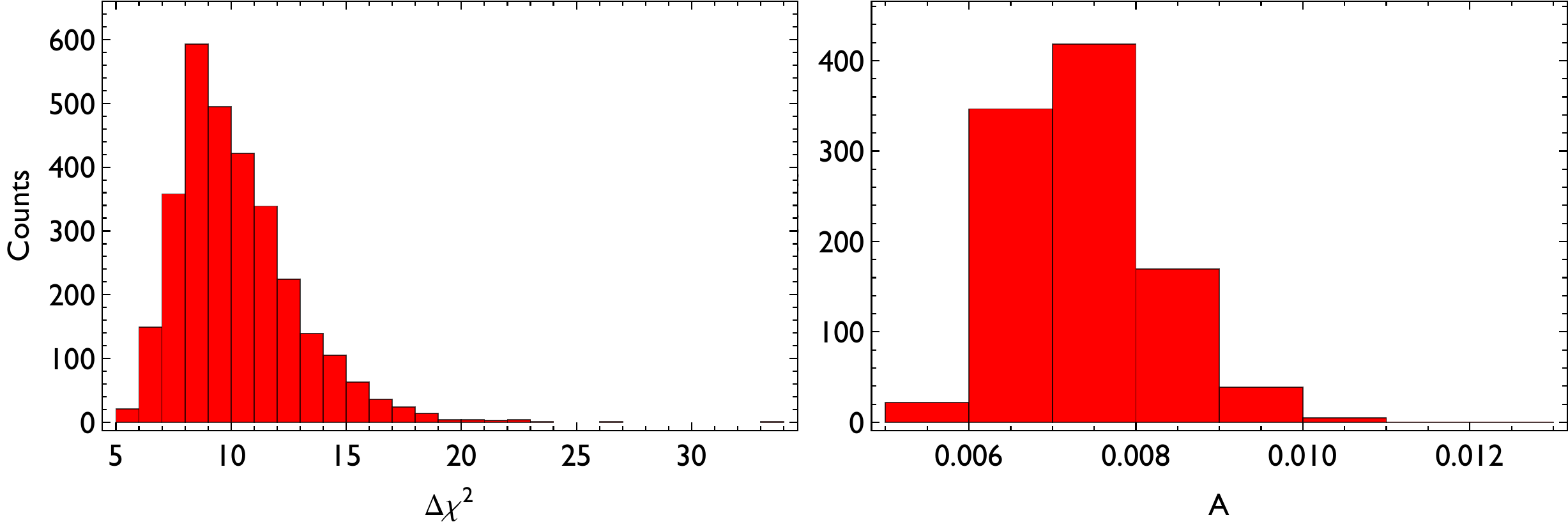} 
   \caption{$\Delta \chi^2$ distribution for 5000 null simulations (right) and distribution of associated best-fit amplitudes (projected). The noise can account for improvement up to $\Delta \chi^2\sim 30$ . }
   \label{fig:distribution}
\end{figure}

We then applied this to the WMAP9 data \cite{Meerburg2014a} and Planck data \cite{Meerburg2014b}. Fig. \ref{fig:multinest} shows the resulting best-fit contour and the marginalized contour for the frequency parameter $\omega_1$ for log spaced oscillations. There are several tentative hints of a significantly improved fit. The results shown are for the Planck data, which contain many more possible features at low frequency than WMAP9 (not shown). We investigated the highest peaks in more detail, by checking the improvement as a function of multipole $\ell$ which strongly suggested that the low frequency features are a consequence of an anomaly in the Planck 217 GHz channel around $\ell \sim 1800$. For linear oscillations the improvements are smaller, but we found no specific multipole that caused them. We did however ran several thousand simulations with a null signal, and applied a simple grid based search to estimate how much the noise can contribute to any given signal. This analysis showed that the distribution of $\Delta \chi^2$, is not a 3 parameter $\Delta \chi^2$ distribution, but instead a distribution that peaks around $\Delta \chi^2 \sim 10$, and $\Delta \chi^2\sim 27$ is the 3 sigma limit (see Fig. \ref{fig:distribution}). Hence, for highly oscillating features, it is very hard to say if one is simply fitting the noise, or an actual residual coming from the primordial power spectrum.

%

\section{Multinest}\label{sec:multinest}
In our analysis we used a MH MCMC for each frequency because a MH sampling is not fit to search a high irregular likelihood. The frequency parameter causes this irregularity and in order to obtain convergence, one needs to fix the frequency. As a result, the time needed to reconstruct the best-fit contours depends on the resolution in frequency space. In our work we considered about 1000 steps. Although these can run a in parallel, this is far from optimal and requires a substantial amount of CPU hours. The nested sampler MULTINEST \cite{Feroz2009,Feroz2013} is a much better sampler in case of irregular posteriors. We recently modified our code to work with multinest. We still precompute the perturbed $C^u_{\ell}$ for many frequencies, but we added a spline routine that allows one to continuously sample through frequency space. The results of the search for log-spaced oscillations is shown in Fig. \ref{fig:multinest}. For reference, one should compare this to Fig. 19 in \cite{PlanckInflation2013}. The advantages of our method compared to the result shown in that figure are twofold. First, we can vary all cosmological parameters (as opposed to just 3 in that particular work), which leads to slightly bigger improvements. Second, we can run up to high frequencies with only limited computational time increase. The total time to reproduce the posterior as shown in Fig. \ref{fig:multinest} is  192 CPU hours (16 hours x 12 cores). We plan to make the code publicly available \footnote{\url{http://www.astro.princeton.edu/~meerburg/coding}.}. 

\section{BICEP}\label{sec:BICEP}
The BICEP detection of B-mode polarization, if primordial, increases tension between the TT data (Planck) and the $\Lambda$CDM + $r$ concordance model. If the foregrounds are minimal (see \cite{2014arXiv1405.7351F} for a discussion), then the large value of $r$ villages the {\it Lyth} bound \cite{Lyth1999,EastherKinney2006}, and indicates the field responsible for inflation, underwent  super Planckian displacement. Although a violation of the bound is a not an actual physical bound that can never be violated, it does pose a concern from a model building perspective, which usually relies on integrating out UV physics and working in an effective framework. When the {\it Lyth} bound is violated, it requires a UV complete theory to fully understand the mechanism of inflation. One working example is axion-monodromy inflation, in which a shift symmetry naturally produces a super Planckian displacements This model also produces features and we investigated if such a feature can lead to an improved fit. Such a feature should be a very long wavelength feature, complementary to the highly oscillating features investigated above. At these long wavelengths, one expects large degeneracies since an oscillation effectively causes a particular rescaling of the amplitude and scale dependence of the power spectrum. Hence, it is necessary to vary all parameters non-perturbatively. In order to avoid getting stuck in a local minimum of the posterior when running through frequency space, we considered a very narrow prior on the frequency, with $10^{-2} \leq \omega \leq 2$ for the log spaced oscillation template of Eq. \ref{eq:powerspectra1}. We also included oscillations in the BB power spectrum for completeness. Our analysis included Planck, BICEP2, ACT, SPT and WMAP polarization data. Quite interestingly, we find an improved fit with $\Delta \chi^2\sim 11$, which has significant modification for $\ell <100$, and no modulations on small scales; the long wavelength oscillation is compensated by a large tilt $n_s\sim1$ to render the spectrum equivalent to scale dependent spectrum with a tilt $n_s=0.96$ on small scales. We explore the potential of such a feature in Ref. \cite{Meerburg2014}.

\section{Conclusions}\label{sec:conclusions}

We reviewed a newly developed technique to look for highly oscillatory features in the CMB power spectrum. This method allows one to vary all cosmological parameters, which has typically been limited by the slow computation of the transfer functions. Applying the method to mock data recovers the mock signal. When applied to the Planck CMB data, we found several tentative hints, which we further investigated. We showed that cosmic variance + noise can account for these findings, as over a sample of 5000 Universes one expects a $\Delta \chi^2 \sim 10$ within the prior volume investigated. For log spaced oscillations the improvement appears to be entirely due a feature in the 217GHz map at $\ell \sim 1800$. We have recently implemented our code in the MULTINEST sampler, which allowed us to recover previous results in only $192$ CPU hours. We briefly discussed the possibility of a long wavelength feature to release some of the tension between the data and the $\Lambda$CDM +$r$ concordance cosmology. We find that there exists a specific feature that can improve the fit significantly. We discuss these findings in Ref.~\cite{Meerburg2014}.

\section*{Acknowledgments}

P.D.M would like to thank the organizers of the Moriond  conference  for organizing an excellent conference. P.D.M. was supported by the John Templeton Foundation grant number 37426.

\section*{References}

\bibliography{moriond_paper}

%
%
%
%

\end{document}